\documentclass[RNAAS]{aastex631}

\hypersetup{linkcolor=blue,citecolor=blue,filecolor=blue,urlcolor=blue}

\received{August 17, 2024}

\submitjournal{RNAAS}

\shorttitle{A false positive transit candidate for EPIC\,211101996 from K2 and TESS}
\shortauthors{Heller et al.}

\begin{document}

\title{A false positive transit candidate for EPIC\,211101996 from K2 and TESS data identified as background eclipsing binary Gaia\,DR3\,66767847894609792}

\correspondingauthor{Ren{\'e} Heller}
\email{heller@mps.mpg.de}

\author[0000-0002-9831-0984]{Ren{\'e} Heller}
\affiliation{Max Planck Institute for Solar System Research, Justus-von-Liebig-Weg 3, 37077 G{\"o}ttingen, Germany}

\author[0009-0009-6573-2425]{Milena Hüschen}
\affiliation{Institut f{\"u}r Astrophysik und Geophysik, Georg-August-Universit{\"a}t G{\"o}ttingen, Friedrich-Hund-Platz 1, 37077 G{\"o}ttingen, Germany}

\author[0000-0001-8935-2472]{Jan-Vincent Harre}
\affiliation{Institute of Planetary Research, German Aerospace Center (DLR), Rutherfordstrasse 2, 12489 Berlin, Germany}

\author[0000-0001-6187-5941]{Stefan Dreizler}
\affiliation{Institut f{\"u}r Astrophysik und Geophysik, Georg-August-Universit{\"a}t G{\"o}ttingen, Friedrich-Hund-Platz 1, 37077 G{\"o}ttingen, Germany}

\begin{abstract}
\noindent
Transiting planets around young stars are hard to find due to the enhanced stellar activity. Only a few transiting planets have been detected around stars younger than 100 Myr. We initially detected a transit-like signal in the K2 light curve of a very cool M dwarf star (EPIC\,211101996) in the Pleiades open cluster, with an estimated age of about 100 Myr. Our detailed analysis of the per-pixel light curves, detrending with the W{\={o}}tan software and transit search with the Transit Least Squares algorithm showed that the source of the signal is a contaminant source (Gaia\,DR3\,66767847894609792) $20\arcsec$ west of the target. The V-like shape of its phase-folded light curve and eclipse depth of $\sim15$\,\% suggest that it is a grazing eclipsing binary. The contaminant has hitherto been listed as a single star, which we now identify as an eclipsing stellar binary with a period of about 6\,days.
\end{abstract}


\section{Introduction}
\label{sec:intro}

\noindent
In their search for transiting planets in 702 light curves of ultra-cool dwarf stars (and possibly brown dwarfs) in K2 data, \citet{2020A&A...641A.170S} did not find any new planets in addition to the previously known TRAPPIST-1 system. In their search, they used the Box Least Squares (BLS) algorithm \citep{2002A&A...391..369K}, which has been shown to be less sensitive to shallow transits than the Transit Least Squares (TLS) algorithm \citep{2019A&A...623A..39H,2019A&A...627A..66H}. We thus repeated the search for transiting objects around those 702 ultra-cool targets, hoping to find small planets that had previously been missed.

\section{Methods}
\label{sec:methods}

\noindent
We took the K2 light curves that had been generated by \citet{2020A&A...641A.170S} from K2 target pixel files using their custom pipeline for flux extraction, systematic noise removal, and Gaussian progress regression \citep{vizier:J/A+A/641/A170}. Then we applied TLS to search for transits, the total computing time of which amounted to just a few hours on a standard laptop. TLS detected a new transit-like signal in the K2 light curve of EPIC\,211101996 (TIC\,35155735) with a period of $5.98796\,(\pm0.01094)$\,d and a depth of about 3.6\,\%. The signal detection efficiency (SDE) of 47.9 exceeded the commonly used threshold value for a detection of 7 by a lot.

Using spectral fitting of the Gaia and 2MASS photometry, \citet{2020A&A...641A.170S} estimate a mass of $0.1\,M_\odot$, a radius of $0.127\,R_\odot$, an effective temperature of 2850\,K, and a spectral type of M6V for EPIC\,211101996. The planet-to-star radius ratio that we initially measured implied a planet radius of about $2.6\,R_\oplus$ for our new transit candidate. The sky position of EPIC\,211101996 and its parallax of 7.36\,mas from Gaia Data Release 3 (DR3) \citep{2023A&A...674A...1G}, equivalent to a distance of about 136\,pc, strongly suggest that it is a member of the Pleiades open star cluster, which has an estimated age of just about 100\,Myr \citep{1996ApJ...458..600B,1998ApJ...497..253U}.

For our vetting of the transiting planet candidate, we used the nominal K2 light curve generated by the EVEREST v2.0 pipeline \citep{2018AJ....156...99L} and applied various detrending methods provided by the W{\={o}}tan software package \citep{2019AJ....158..143H}. None of our detrending methods resulted in a detection of the signal that we found in the light curve provided by \citet{2020A&A...641A.170S}. We also tested other corrections of the K2 light curve from systematic effects such as k2sc, k2sff, k2varcat, polar and ktwo, which did not result in a confirmation of the signal.

From August to November 2021, new TESS observations of EPIC\,211101996 were obtained in Sectors 42, 43, and 44. The TESS data validation report summary showed a clear transit signal with a period of $11.97649\,(\pm\,0.00578)$\,d, which is 2.000095 times the period that we found in the K2 data. More suspiciously, however, the transit depth was now a whopping 25\,\%, suggesting a radius for the transiting object of 50\,\% of the primary radius, and thus a possible stellar binary rather than a transiting planet.

\begin{figure}[ht!]
\centering
\includegraphics[width=1.0\linewidth]{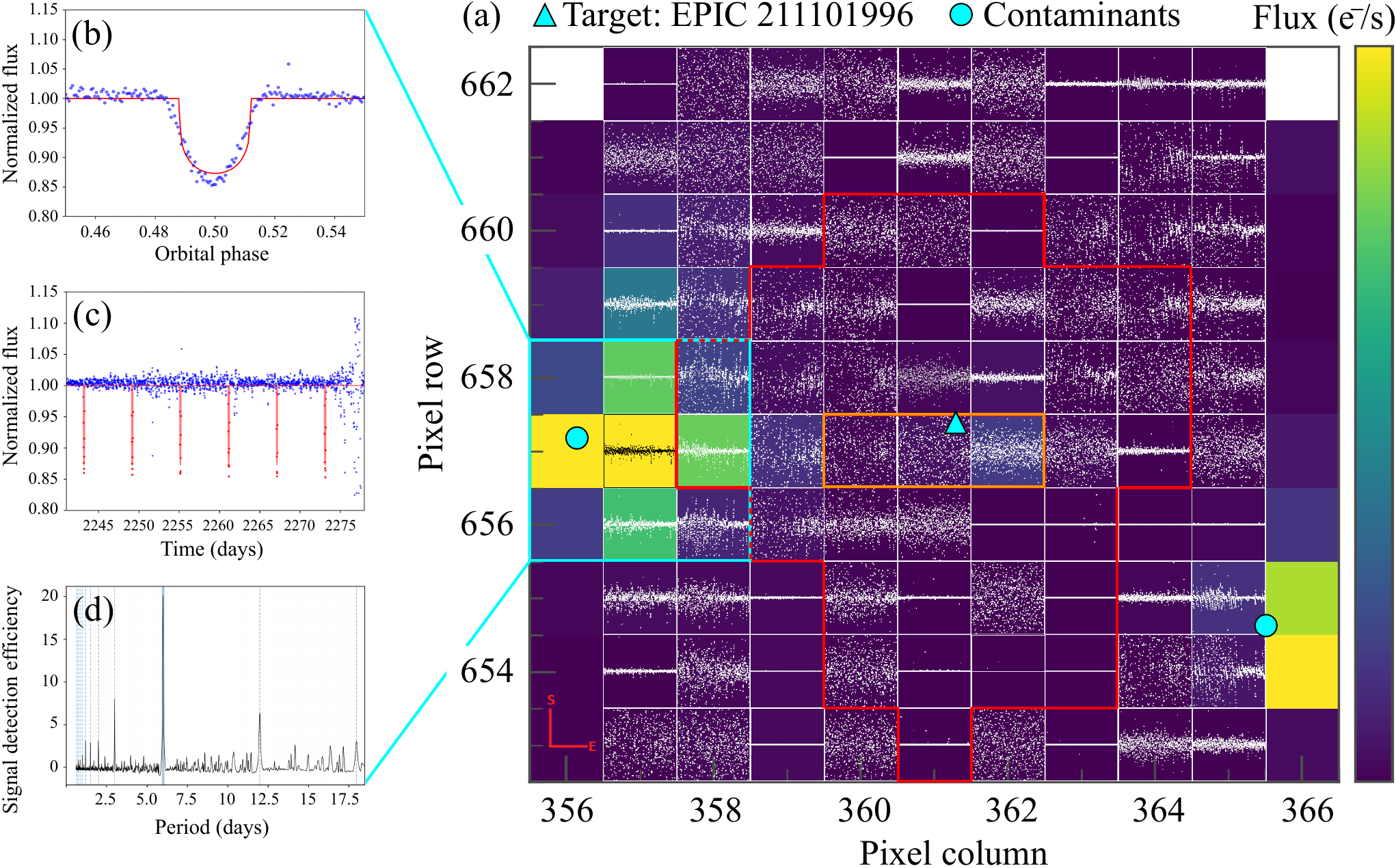}
\caption{(a) K2 target pixel file for EPIC\,211101996 extracted with the Lightkurve software \citep{2018ascl.soft12013L}. Each pixel contains a plot of the per-pixel light curve for 2247\,d--2269\,d Barycentric TESS Julian Date (TBJD). The thin orange line in the center denotes the nominal K2 aperture. The red line outlines the nominal aperture used by EVEREST. The light blue line around the $3 \times 3$ pixels at the left (west) shows the aperture mask that we used to localize the periodic signal around the contaminant Gaia\,DR3\,66767847894609792. (b) TLS output of the phase-folded light curve for the signal identified around the contaminant. (c) Detrended light curve of the contaminant with in-eclipse data points highlighted in red. (d) SDE periodogram of the contaminant with a strong peak at 5.98796\,d.}
\label{fig:pixel}
\end{figure}

\section{Results}
\label{sec:results}

\noindent
We tested multiple aperture masks around EPIC\,211101996 using EVEREST to extract the K2 light curves, then detrended them with Tukey's biweight filter implemented in W{\={o}}tan and searched for the origin of the transit-like signal with TLS. We identified the contaminant Gaia\,DR3\,66767847894609792, which is about $20\arcsec$ west of the target, as the source of the signal (Fig.~\ref{fig:pixel}). TLS found an SDE value of $\sim20$ for the light curve that we extracted (see Fig.~\ref{fig:pixel}c), which is substantially lower than the signal that we originally detected in the light curve provided by \citet{2020A&A...641A.170S}.

Gaia\,DR3\,66767847894609792 has hitherto been classified as a single stellar object in Gaia DR3 with a parallax of $0.3967\,(\pm\,0.0428)$\,mas. The resulting distance of $2520_{-245}^{+305}$\,pc and its proper motion $\mu_{\alpha}=2.454\,(\pm\,0.045)$\,mas/yr, $\mu_{\delta}=-2.896\,(\pm\,0.040)$\,mas/yr suggest that the astrophysical source of the transit-like signal is in a chance alignment far behind the Pleiades cluster.

The V-like shape and ${\sim}15$\,\% depth of the phase-folded light curve (see Fig.~\ref{fig:pixel}b) plus visual indication of a shallow secondary eclipse indicate an eclipsing stellar binary as the source of the signal. Moreover, the Markov chain Monte Carlo multiple-star classifier from Gaia DR3 (not shown) exhibits a bimodal posterior distribution with primary stellar parameters $T_{\rm eff,1}\,{\sim}\,5550$\,K, $\log(g_{\rm 1})\,{\sim}\,4.4$, secondary stellar parameters $T_{\rm eff,2}\,\sim\,5100$\,K, $\log(g_{\rm 2})\,\sim\,4.75$, a combined metallicity of $[{\rm M/H}]\,{\sim}\,-0.7$\,dex, and a distance of about 1730\,pc. This revised distance estimate is based on low-resolution BP/RP spectra and the parallax estimate from Gaia \citep{2023A&A...674A...1G}, still suggesting a position far behind the Pleiades from our perspective.

The TESS Science Processing Operations Center (SPOC) pipeline derived an alias of the true orbital period due to gaps in the observations that masked several transits. SPOC also falsely attributed the signal to EPIC\,211101996, partly because the TESS pixels are much larger than the K2 pixels. A detailed comparison of the TESS and K2 aperture masks and the resulting depths of the eclipse signals was beyond the scope of our study.

\section{Conclusions}
\label{sec:Conclusions}

\noindent
The transit-like periodic signal in the K2 and TESS light curves of EPIC\,211101996 (TIC\,35155735) is a false positive caused by a background contaminant, which we identify as an eclipsing stellar binary. The source of the signal is Gaia\,DR3\,66767847894609792, which should be considered as a stellar binary rather than a single source.

\begin{acknowledgments}
RH acknowledges support from the German Aerospace Agency (Deutsches Zentrum f\"ur Luft- und Raumfahrt) under PLATO Data Center grant 50OO1501. JVH is funded by the DFG priority programme SPP 1992 ``Exploring the Diversity of Extrasolar Planets (SM 486/2-1)''. The authors thank Matthias Ammler-von Eiff for helpful comments.
\end{acknowledgments}

\vspace{5mm}
\facilities{Kepler (K2), TESS, Gaia}

\software{TLS \citep{2019A&A...623A..39H},
                W{\={o}}tan \citep{2019AJ....158..143H},
                lightkurve \citep{2018ascl.soft12013L}
               }

\bibliography{new.ms}{}
\bibliographystyle{aasjournal}

\end{document}